\documentstyle[preprint,aps,epsf]{revtex}
\def\baselinestretch{1.5}

\newcommand{\be}[1]{\begin{equation} \label{(#1)}}
\newcommand{\ee}{\end{equation}}
\newcommand{\ba}[1]{\begin{eqnarray} \label{(#1)}}
\newcommand{\ea}{\end{eqnarray}}

\def \znbb {$0\nu\beta\beta~$}

\def\lsim{\mathrel{\vcenter{\hbox{$<$}\nointerlineskip\hbox{$\sim$}}}}
\def\gsim{\mathrel{\vcenter{\hbox{$>$}\nointerlineskip\hbox{$\sim$}}}}
\def\egzk{E_{\rm GZK}}

\def\nue{\nu_e}

\def\numu{\nu_{\mu}}
\def\nutau{\nu_{\tau}}
\def\nuebar{\bar{\nu}_e}

\def\dmsq{\delta m^2}
\def\dmatm{\delta m^2_{\rm atm}}
\def\dmsun{\delta m^2_{\rm sun}}

\begin{document}
\preprint{\vbox{\hbox{VAND--TH--01--1}
\hbox{hep-ph/0101091}
\hbox{January 9, 2001}
}}
\title{Absolute Neutrino Mass Determination}
\author{H. P\"as
\footnote[1]{E-mail: heinrich.paes@vanderbilt.edu} 
and 
T.J. Weiler
\footnote[2]{E-mail: tom.weiler@vanderbilt.edu}
}
\address{Department of Physics and Astronomy\\ Vanderbilt University\\
Nashville, TN 37235, USA}

\maketitle

\begin{abstract}
We discuss four approaches to the determination of 
absolute neutrino mass.  These are the measurement of 
the zero-neutrino double beta decay rate,
of the tritium decay end-point spectrum, of the cosmic ray 
spectrum above the GZK cutoff (in the Z-burst model),
and the cosmological measurement of the power spectrum governing the 
CMB and large scale structure.
The first two approaches are sensitive to the mass eigenstates coupling to 
the electron neutrino, whereas the latter two are sensitive to the heavy
component of the cosmic neutrino background.
All mass eigenstates are related by the $\dmsq$'s inferred from 
neutrino oscillation data.  Consequently, the potential for
absolute  mass determination of each of the four approaches is 
correlated with the other three, in ways that we point out.\\
\\
\noindent
PACS: 14.60.Pq, 23.40.Bw, 96.40.Tv, 98.80.Es.

\end{abstract} 

\newpage

\section{Introduction}

An ongoing experimental effort of high importance 
is the determination of the neutrino
mass eigenvalues.
The absolute scale of neutrino masses, 
a crucial datum for reconstructing physics beyond the Standard
Model, is unknown.
Presently, upper bounds on the absolute neutrino mass are provided by 
the tritium decay end-point spectral measurement, by cosmology, and
by zero-neutrino double beta decay (\znbb).

The present tritium decay upper bound on each \cite{bww98} of
the three neutrino mass eigenstates 
is 2.3 eV (95 \% C.L.) \cite{tritium}. 
An upper bound from cosmological structure formation is more stringent
but also more model-dependent.
For three degenerate neutrino masses, the 
constraint on the individual neutrino mass eigenstates is 
$m_j < 1.8$~eV for large $\Omega_m$,
and  $m_j < 0.6$~eV for $\Omega_m \sim 0.3$ \cite{primack00};
$\Omega_m$ is the matter fraction of the closure density.
The present \znbb\ upper limit on the $ee$ element $m_{ee}$ 
of the flavor-basis Majorana neutrino mass matrix is 0.27~eV \cite{bbdata}.
Fortunately for \znbb\ searches, 
models which generate small neutrino masses overwhelmingly favor 
Majorana neutrinos over Dirac neutrinos \cite{numassrevs}
(but see \cite{Smirnov98} 
for a small Dirac masses generated by brane-bulk interactions).

To determine an absolute neutrino mass below 1~eV is a true
experimental challenge.
The three approaches just mentioned 
have the potential to accomplish the task.  
Anticipated improvements in these approaches are\\
(i)  larger versions of the tritium end-point distortion measurements;\\
(ii) the comparison of more-precisely determined early-Universe 
temperature perturbations (MAP \cite{MAP} and PLANCK \cite{Planck} experiments)
to the present-day large-scale structure distributions of 
matter (to be measured by SDSS \cite{SDSS} and 2dF \cite{2dF}); and \\
(iii) larger \znbb\ experiments (GENIUS and EXO are proposed).\\
In addition there is a fourth possibility: \\
(iv) the extreme-energy cosmic-ray experiments 
(AGASA \cite{agasa}, HiRes \cite{hires}, Auger \cite{auger}, Telescope Array
\cite{telarr}, EUSO/OWL \cite{EUSO/OWL})
in the context of the recently emphasized Z-burst model
\cite{wei82,Zburst}.\\
Still another approach to neutrino mass determination,
measuring the arrival-time profile of neutrinos
from supernovae, seems not quite capable of breaking the sub-eV
barrier \cite{BBM00}.

The Z-burst and cosmic structure measurements 
are sensitive to the heavier neutrino masses 
(and the cosmological neutrino background),
while the tritium and \znbb\ experiments 
are sensitive to different linear combinations of
whichever masses 
are coupled to $\nue$.
Neutrino oscillation interpretations of solar, atmospheric, and
LSND data produce nonzero values for neutrino mass-squared differences,
and so relate all neutrino masses.
Accordingly, the expectations of the four approaches listed above to 
absolute neutrino mass determination are related.  Any positive finding
in one approach requires concordance in the other three.
It is the purpose of this work to reveal the relations among the
reaches of the four approaches.

We begin with a discussion of the neutrino mass-relations inferred
from oscillation interpretations of recent data.
We will conservatively consider a three-neutrino Universe,
omitting the uncorroborated 
data from the LSND experiment;
in the future, the miniBooNE experiment at Fermilab will rule on
the validity of the LSND measurement.
Specifically, we label the mass eigenstates as $m_3> m_2> m_1$,
and denote the mass-squared differences as 
\be{}
 m_3^2 =m_1^2 + \dmatm\,,\quad {\rm and} \quad m_2^2=m_1^2 + \dmsun\,
\label{onepar}
\ee
with $\dmatm$ and $\dmsun$ positive.
Oscillations are directly sensitive to these
nonzero neutrino mass--squared differences.
The alternative splitting (``inverted hierarchy'') 
with two heavy
states and a single light state is
discussed briefly in the conclusions section; 
it is disfavored according to a recent analysis \cite{inv}
of the neutrino spectrum from SN1987A, unless 
the mixing element $U_{e1}$ is large.

The solar and 
atmospheric neutrino oscillation interpretations, 
and the nonobservation of short-baseline $\nue$ disappearance in 
the CHOOZ experiment, provide valuable
information on the $\dmsq$'s and mixing angles. 
The most recent global data-analysis in a three neutrino framework
yields the following favored regions \cite{gon99}:\footnote
{It has become customary to express a mixing angle 
sensitive to matter effects, such as the solar angle, 
as $\tan\theta$ rather than $\sin2\theta$
to account for the octant of the ``dark side'',
$\pi/4 <\theta\leq\pi/2$.}
\begin{itemize}

\item
{\bf Solar neutrino data} favor $\nu_e-\nu_{\not e}$ 
oscillations within the \\
large mixing-angle {\bf (LMA)} MSW solution:\\
$\dmsun=3\times 10^{-5}$~eV$^2$, 
with a 90\% C.L. of $(1-10)\times 10^{-5}$~eV;\\
$\tan^2 \theta_{\rm sun}= 0.5$,
with a 90\% C.L. of (0.2-0.6). \\
\\
Also allowed at 90 \% C.L. is a small region in the \\
{\bf LOW-QVO} (quasi-vacuum oscillation) regime: \\
$\dmsun= 10^{-7}$~eV$^2$,\\
$\tan^2 \theta_{\rm sun}=(0.6-0.8)$.\\
\\
The small mixing-angle {\bf (SMA)} MSW solution at\\ 
$\dmsun= (4-9) \cdot 10^{-5}$~eV$^2$,\\
$\tan^2 \theta_{\rm sun}=(0.2-1)\cdot 10^{-3}$\\
is disfavored at 90 \% C.L. but viable at 95 \% C.L.

\item
{\bf Atmospheric neutrino data} are explained by 
{\bf maximal} $\nu_{\mu}-\nu_{\tau}$ 
oscillations with: \\
$\dmatm = 3\times 10^{-3}$~eV$^2$,
and a 90\% C.L. of $(1.6-5)\times 10^{-3} {\rm eV}^2$;\\
$\sin^2 2\theta_{\rm atm} > 0.85$.

\end{itemize}
It should be stressed also that at larger C.L. 
the large-angle solution for solar neutrinos can extend over
nearly the entire region from 
$\dmsun =10^{-10}$~eV$^2$ up to $\dmsun =10^{-3}$~eV$^2$.
Also, data from Supernova1987A have recently been re-analyzed in the context 
of the various solar neutrino solutions.  The result is that
the LOW-QVO solutions are disfavored at $4\sigma$ compared to the
SMA and LMA solutions \cite{SN87a} 
over most of the supernova parameter space.

The mass-squared differences inferred from solar and atmospheric measurements
imply lower bounds on the masses
$m_3$ and $m_2$.  The atmospheric bound is  
$m_3\ge\sqrt{\dmatm}\sim 0.05$~eV,
which is encouraging for mass-sensitive experiments.
The relation among the three masses enforced by the 
oscillation interpretation of solar and atmospheric data 
is plotted is Fig.\ 1.
Also shown in the figure are the present tritium and cosmological
upper bounds on absolute neutrino mass.
The mass-squared differences inferred from data 
show a definite hierarchy:
$\dmsun\ll \dmatm$ by probably a factor of 30 or more.
As seen in Fig.\ 1,
this may or may not imply a mass hierarchy.
If $m_1\gg \sqrt{\dmatm}\sim 0.05\,{\rm eV}$,
then all three neutrino masses are nearly degenerate,
while if  $m_1\lsim \sqrt{\dmatm}\sim 0.05\,{\rm eV}$,
then the three masses are not degenerate.
The degenerate possibility has 
been preferred in cold+hot dark-matter models to account for
observed large-scale structures.
However, the need for the hot component is mitigated by the 
cosmological constant introduced
to explain high red-shift Type Ia supernovae observations. 

If $m_1\gg \sqrt{\dmsun}\sim 0.003\,{\rm eV}$,
then the two lightest neutrino masses $m_1$ and $m_2$ 
are nearly degenerate.
With the exception of a futuristic 10 ton 
version of GENIUS,
the reach of the four approaches considered 
in this work does not extend down to as low as 0.003~eV.
Accordingly, in what follows 
we take $m_1$ and $m_2$ to be degenerate.

We return to the four approaches to absolute neutrino-mass determination.
Because the relevance of 
extreme-energy cosmic rays (EECRs) to neutrino mass determination 
via the Z-burst model is the least known of the approaches,
and because data already exist which
in the context of the model implicate an absolute 
neutrino mass (in the range 0.1 to 1.0 eV),
we consider the Z-burst approach first. 
The model is speculative.  
However, if it is validated
as the explanation of EECR puzzles, the payoff is big.
Not only is the absolute mass of the neutrino revealed, 
but also the existence
of the cosmic neutrino background (CNB) liberated one second
after the Big Bang.

\section{The Z-burst model for EECR's}

It was expected that the EECR
primaries would be protons from outside the galaxy, produced in
Nature's most extreme environments such as the tori or radio hot spots 
of active galactic nuclei (AGN).
Indeed, cosmic ray data show a spectral flattening just below
$10^{19}$~eV
which can be interpreted as a new extragalactic component overtaking
the lower energy galactic component;
the energy of the break correlates well with the onset of a Larmor radius 
for protons too large to be contained by the Galactic magnetic field.
It was further expected that the extragalactic spectrum 
would reveal an end at the 
Greisen-Kuzmin-Zatsepin (GZK) cutoff energy of
$\egzk \sim 5\times 10^{19}$~eV. 
The origin of the GZK cutoff is the degradation of nucleon energy by the 
resonant scattering process $N+\gamma_{2.7K}\rightarrow \Delta^*
\rightarrow N+ \pi$ when the nucleon is above the resonant threshold $\egzk$ 
for $\Delta^*$ production.  The concomitant energy-loss factor is
$\sim (0.8)^{D/6 {\rm Mpc}}$ for a nucleon traversing a distance $D$. 
Since no AGN-like sources are known to exist within 100
Mpc of earth, the energy requirement for a proton arriving at earth with a
super-GZK energy is unrealistically high. 
Nevertheless, to date more than twenty events with energies 
at and above $10^{20}$~eV have been observed \cite{EECRdata}. \\
The spectral break just below $\sim 10^{19}$~eV and the super-GZK
events from the AGASA experiment are displayed in Fig.\ 2.

Several solutions have been proposed
for the origin of these EECRs,
ranging from unseen Zevatron accelerators (1~ZeV~$=10^{21}$~eV) 
and decaying supermassive particles and topological 
defects in the Galactic vicinity, 
to exotic primaries, exotic new interactions, and even exotic breakdown
of conventional physical laws \cite{EECRreviews}.
A rather conservative and economical scenario involves cosmic ray 
neutrinos scattering resonantly on the cosmic neutrino background (CNB) 
predicted by Standard Cosmology, 
to produce Z-bosons \cite{Zburst}. 
These Z-bosons in turn decay to produce a highly boosted ``Z-burst'',
containing on average twenty photons and two nucleons above $\egzk$
(see Fig.\ 3).
The photons and nucleons from Z-bursts produced within 50 to 100 Mpc
of earth can reach earth with enough energy to initiate the 
air-showers observed at $\sim 10^{20}$~eV.

The energy of the neutrino annihilating at the peak of the Z-pole is
\be{}
E_{\nu_j}^R=\frac{M_Z^2}{2 m_j}=4\,(eV/m_j)\,{\rm ZeV}.
\ee
The resonant-energy width is narrow, 
reflecting the narrow width of the Z-boson: at FWHM 
$\Delta E_R/E_R \sim\Gamma_Z/M_Z = 3\%$.
The mean energies of the $\sim 2$~baryons and $\sim 20$~photons
produced in the Z decay are easily estimated.
Distributing the Z-burst energy among the mean multiplicity 
of 30 secondaries in Z-decay \cite{RPP},
one has 
\be{}
 \langle E_p \rangle \sim \frac{E_R}{30} 
\sim 1.3 \left(\frac{{\rm eV}}{m_j}\right)\times 10^{20}{\rm eV}\,.
\ee
The photon energy is further reduced by an additional factor of 2 
to account for their origin in two-body $\pi^0$ decay:
\be{}
 \langle E_{\gamma} \rangle \sim \frac{E_R}{60} 
\sim 0.7 \left(\frac{{\rm eV}}{m_j}\right)\times 10^{20}{\rm eV}\,.
\ee
Even allowing for energy fluctuations about mean values, 
it is clear that in the Z-burst model the relevant
neutrino mass cannot exceed $\sim 1$~eV.
On the other hand, the neutrino mass cannot be too light
of the predicted primary energies will exceed the observed
event energies.\footnote
{Also, the neutrino mass cannot be too small without pushing
the primary neutrino flux to unattractively higher energies.}
In this way,
one obtains a rough lower limit on the
neutrino mass of $\sim 0.1$~eV for the Z-burst model,
when allowance is made for an order of magnitude energy-loss 
for those secondaries traversing 50 to 100 Mpc. 

The necessary conditions for the viability of this 
model are a sufficient flux of neutrinos at $\gsim 10^{21}$ eV 
and a neutrino mass scale of the order $0.1-1$ eV \cite{wei82,Zburst}.
The first condition seems challenging \cite{nuFlux},
while the second is quite natural in view of the recent oscillation data
(see Fig.\ 1).

It is worth remarking that the cosmic fluxes 
of the three neutrino mass-eigenstates 
are virtually guaranteed to be nearly equal 
as a result of the $\numu - \nutau$ near-maximal
mixing observed in atmospheric data.
This comes about as follows:  
For extragalactic neutrinos
produced in $\pi^{\pm}$ decay, the original flavor ratio 
$\nue:\numu:\nutau\sim 1:2:0$ oscillates to $\sim 1:1:1$;
for more exotic neutrino production from, e.g., string cusps,
a flavor-neutral ratio of $1:1:1$ may be expected at the source.
For both cases, an equal population of flavor states results 
for the cosmic flux.
It then follows from unitarity of 
the mixing matrix that
there is also an equal population of mass states in the flux.\footnote
{
The mass basis is the more relevant basis for annihilation on the 
nonrelativistic relic neutrinos;
it is also the more physical basis when it is realized that flavor states
traveling cosmic distances (the flux) or existing for cosmic ages 
(the CNB) will have decohered into mass states.
}
An equal population of mass states is also expected among the thermally
produced CNB.  
The equal population of mass states in flux and CNB 
has interesting consequences.
It follows that the relative Z-burst rate at each of the three 
resonant energies is given by the total neutrino 
flux value $F_\nu (E^R)$ 
at each resonant energy.
If the neutrinos are mass degenerate, then a further consequence is that
the Z-burst rate at $E_R$ is three times what it would be 
without degeneracy. This ameliorates slightly the formidable
flux requirement.

The viability of the Z-burst model is enhanced if the CNB neutrinos 
cluster in our matter-rich vicinity of the universe.
The main constraints on clustering are two-fold.
For very large scales, the Universe is too young to have experienced
significant infall of matter.  
For smaller scales, the Pauli blocking of identical
neutrinos sets a limit on density enhancement.
As a crude
estimate of Pauli blocking, one may use the zero temperature Fermi gas as a
model of the gravitationally bound neutrinos. Requiring that the Fermi
momentum of the neutrinos does not exceed mass times the virial velocity 
$\sigma\sim\sqrt{MG/L}$ within the cluster of mass $M$ and size $L$, 
one gets \cite{TG80s}
\be{}
\frac{n_{\nu_j}}{54\,{\rm cm}^{-3}}\lsim 
10^3 \left(\frac{m_j}{{\rm eV}}\right)^3 
\left(\frac{\sigma}{200{\rm km/s}}\right)^3\,.
\ee
The virial velocity within our Galactic halo 
is a couple hundred km/sec.
Thus it appears that Pauli blocking allows significant clustering on the
scale of our Galactic halo only if $m_j \gsim 0.5$~eV.
An indicator of the neutrino mass
sufficient to allow a 100-fold increase in 
the Galactic halo density of the CNB is shown on Fig.\ 4.

For rich clusters of galaxies, the virial velocities 
are a thousand km/s or more.  Thus, Pauli blocking does not exclude
significant clustering on scales of tens of Mpc for $m_j\gsim 0.1$~eV.
However, on these large scales, the infall of matter integrated 
to the present time is probably insufficient to effect significant clustering.

\section{Tritium decay end-point limits}

In tritium decay, the larger the mass states comprising $\nuebar$,
the smaller is the Q-value of the decay.
The manifestation of neutrino mass is a reduction of phase space
for the produced electron at the high energy end of its spectrum.
An expansion of the decay rate formula about $m_{\nue}$ leads to
the end point sensitive factor 
\be{}
m^2_{\nu_e}\equiv \sum_j\,|U_{ej}|^2\,m^2_j\,,
\ee
where the sum is over mass states which can kinematically alter
the end-point spectrum.
If the neutrino masses are nearly degenerate,
then unitarity of $U$ leads immediately to a bound on
$\sqrt{m^2_{\nu_e}}=m_3$.
The design of larger tritium decay experiments to reduce 
the present 2.3~eV $m_{\nu_e}$ bound are under discussion;
direct mass limits as low as 0.4~eV, or even 0.2~eV, may be possible
in this type of experiment.

\section{CMB/LSS cosmological limits}

According to Big Bang cosmology, 
the masses of nonrelativistic neutrinos are related to the neutrino 
fraction of closure density by
$\sum_j m_j = 40\,\Omega_{\nu}\,h_{65}^2$~eV,
where $h_{65}$ is the present Hubble parameter in units of 65~km/s/Mpc.
As knowledge of large-scale structure (LSS) formation has evolved,
so have the theoretically preferred values for the hot dark matter (HDM)
component, $\Omega_\nu$.  In fact, the values have declined.
In the once popular HDM cosmology, one 
had $\Omega_\nu \sim 1$ and $m_\nu \sim 10$~eV 
for each of the mass-degenerate neutrinos.
In the cold-hot CHDM cosmology, the cold matter was dominant 
and one had $\Omega_\nu\sim 0.3$ and $m_\nu \sim 4$~eV
for each neutrino mass.
In the currently favored $\Lambda$DM cosmology,
there is scant room left for the neutrino component.
An analysis relating the cosmic microwave background 
(CMB) temperature fluctuations 
and the present-day LSS provides the limit.
The power spectrum of early-Universe density perturbations,
fossilized in the observed CMB fluctuations at the recombination 
epoch  $z_r\sim 1100$,
is processed by gravitational instabilities.
However, 
the free-streaming relativistic 
neutrinos suppress the growth of fluctuations
on scales below the horizon 
(approximately the Hubble size $c/H(z)$) 
until they become nonrelativistic at 
$z\sim m_j/3T_0 \sim 1000\,(m_j/{\rm eV})$.
The result of simulation is a neutrino component
constrained as $\sum_j m_j<5.5$~eV for all values of $\Omega_m$;
and $m_j < 0.6\,(0.9)$~eV for each of three degenerate neutrinos
and for $\Omega_m=0.3\,(0.4)$, all at 95 \% C.L. \cite{primack00}.

The Sloan Digital Sky Survey (SDSS) should measure the
power spectrum of the LSS to $\sim 1\%$ accuracy.
Combining this with the CMB 
measurements expected from the MAP satellite,
one can infer neutrino mass down to \cite{tegmark}
\be{}
\sum m_{\nu} \simeq 0.33~ 
\left(\frac{\Omega_m\,h_{65}^2}{0.3}\frac{3}{\rm N} \right)^{0.8}~ 
{\rm eV}.
\ee
Here N is the number of degenerate neutrinos.
The effect of a single neutrino state on the CMB 
anisotropy in $\Lambda$DM  models has also been 
discussed \cite{lopez}.  A sensitivity for MAP to 2~eV neutrinos 
with temperature data alone, 
and to 0.5~eV with polarization data included is estimated; 
for the PLANCK satellite a sensitivity to 0.5~eV with temperature data
alone and to 0.25~eV with polarization data included is claimed.

Some caution is warranted in the cosmological approach to neutrino mass,
in that the many cosmological parameters may conspire in 
various combinations to yield nearly identical CMB and LSS data.
An assortment of very detailed data may be needed to resolve 
the possible ``cosmic ambiguities''.

\section{Neutrinoless double beta decay}

Neutrinoless double beta decay (\znbb) proceeds via the nuclear reaction
$^{A}_{Z}X~\rightarrow~^A_{Z+2}X~+~2~e^-$.
The rate is a sensitive tool for the
measurement of the absolute mass-scale for Majorana neutrinos \cite{kla00}.
The observable measured in the amplitude of \znbb\ decay
is the $ee$ element of the neutrino mass-matrix in the flavor basis.
Expressed in terms of the mass eigenvalues and 
neutrino mixing-matrix elements, it is 
\be{}
m_{ee}= |\sum_i U_{ei}^2 m_i|\,.
\label{dbeqn}
\ee
A reach as low as $m_{ee}\sim 0.01$~eV seems possible 
with proposed double beta decay projects such as the 1 ton version of
GENIUS \cite{genius} and EXO \cite{exo}. 
This provides a substantial improvement over the current bound,
$m_{ee}< 0.27$~eV.
In the far future,
another order of magnitude in reach 
is available to the 
10 ton version of GENIUS, should it be funded and commissioned.

For masses in the interesting range $\gsim 0.01$~eV, 
the two light mass eigenstates are nearly degenerate and so the 
approximation $m_1 =m_2$ is justified.
Furthermore, the restrictive CHOOZ bound \cite{chooz}, $|U_{e3}|^2 < 0.025$
in the three neutrino model (for $\dmatm \geq 10^{-3} eV^2$), 
allows two further simplifications.  The first is that 
the contribution of the third mass eigenstate 
is nearly decoupled from $m_{ee}$ and so
$U^2_{e3}\,m_3$ may be neglected in the \znbb\ formula.
The second is that the two-neutrino mixing approximation is valid, i.e.
$U_{e1}\approx e^{i\phi_1}\cos\theta_{\rm sun}$ and
$U_{e2}\approx e^{i\phi_2}\sin\theta_{\rm sun}$.
We label by $\phi_{12}$ the relative phase between
$U^2_{e1}\,m_1$ and $U^2_{e2}\,m_2$.
Then, employing the above approximations,
we arrive at a very simplified expression for $m_{ee}$:
\be{}
m^2_{ee}=\left[1-\sin^2 (2\theta_{\rm sun})\,
       \sin^2 \left(\frac{\phi_{12}}{2}\right)\right]\,m^2_1\,.
\label{dbeqn2}
%
\ee
The two CP-conserving values of $\phi_{12}$ are 0 and $\pi$.
These same two values give maximal constructive and destructive
interference of the two dominant terms in eq.\ (\ref{dbeqn}),
which leads to upper and lower bounds for the observable
$m_{ee}$ in terms of a fixed value of $m_1$:
\be{}
\cos (2\theta_{\rm sun})\;m_1 \leq m_{ee} \leq m_1 \,,
\quad {\rm for\;\;fixed}\;\;m_1\,.
\label{dbbnds}
\ee
The upper bound becomes an equality, $m_{ee}=m_1$, 
for any of the solar solutions if $\phi_{12}=0$,
and for the small-angle SMA solution 
($\cos (2\theta_{\rm sun})\approx 1$) 
with any $\phi_{12}$.
The lower bound depends on Nature's value of the mixing angle
in the LMA and LOW-QVO solutions.\footnote
{
Were $\theta_{\rm sun}$ truly maximal at $\pi/4$,
the dominant terms in $m_{ee}$ could cancel,
leaving $m_{ee}$ beyond the reach of foreseeable \znbb\ experiments.
Truly maximal mixing is not favored by fits to the data,
nor by any theory.}
%
A consequence of eq.\ (\ref{dbbnds}) is that for a given
measurement of $m_{ee}$, the corresponding inference of $m_1$ is 
uncertain over the range 
$[m_{ee},\,m_{ee}\,\cos (2\theta_{\rm sun})]$
due to the unknown phase difference $\phi_{12}$.

Knowing the value of $\theta_{\rm sun}$ better will improve
the estimate of the inherent uncertainty in $m_1$.
For the LMA solar solution, the 
forthcoming Kamland experiment should reduce the error in the 
mixing angle $\sin^2 2 \theta_{\rm sun}$ to $\pm 0.1$ \cite{barger00}.
However, it is unlikely that the inherent uncertainty in $m_1$  
can be reduced beyond $(\cos 2\theta_{\rm sun})^{-1}$,
since there is no known way to measure the 
Majorana phase difference $\phi_{12}$.
Ultimately, the inferences made for $m_{ee}$ 
from a positive \znbb\ result 
will also depend on the uncertainty in the charged-current 
nuclear matrix element.
Currently this uncertainty is a factor of 2 to 3.
We ignore it in what follows.

A quantitative discussion of the reach of \znbb\ 
is presented in the next section.

\section{Correlations among approaches}

It is evident that the puzzle of absolute scale of neutrino masses 
connects very different branches of physics, ranging from the sub-eV scale
of \znbb\ and end-point tritium decay, 
to the ZeV scale of EECRs, to the matter fluctuations of the
primordial Universe.  
As mentioned in the introduction,
the Z-burst and cosmic structure measurements 
are sensitive to the heavier neutrino masses,
while the tritium and \znbb\ experiments are sensitive to 
linear combinations of masses (presumably the lighter ones) 
most coupled to $\nue$.
The heavier and lighter masses are related by the 
$\dmsq$'s inferred from oscillation experiments,
which in turn correlates the possible findings of the four
approaches to absolute neutrino-mass determination.
One way to display the correlations among the approaches
is to show the overlap of their respective reaches on a mass plot.
This is done in Fig.\ 4, where we
take \znbb\ as representative of the effort to measure
the lighter neutrino masses, and the Z-burst model as
representative for $m_3$ measurement.
The complementary limits from tritium decay and cosmology 
were already presented in the $m_3 -m_1$ plane of Fig.\ 1.

Shown in Fig.\ 4 is the \znbb-observable $m_{ee}$
predicted for each solar solution, 
as a function of the heaviest neutrino 
mass $m_3$, or, alternatively the Z-burst resonance energy 
$E_R=4\,({\rm eV}/m_3)$~ZeV.
This mass-correlated plot is possible because 
fixing $m_3$ fixes $m_1$ and $m_2$,
as given by eqn.\ (\ref{onepar}).
According to eqn.\ (\ref{dbeqn2}), for each solar solution
there results a band of allowed $m_{ee}$, reflecting 
the uncertainties 
in the relative Majorana phase difference $\phi_{12}$.
The exception is the SMA solution, for which the band collapses to 
a unique relation between $E_R$ and $m_{ee}$,
independent of $\phi_{12}$.
For the large-mixing solutions,
the allowed $m_{ee}$ varies by a factor of 4 between $\pi_{12}=0$
and $\pi_{12}=\pi$ for LMA,
and a factor of $\sim$ 10 for LOW-QVO.

Also shown on the Figure are the recent  
Heidelberg-Moscow (HM) bound on $m_{ee}$, 
and the expected GENIUS/EXO sensitivity to $m_{ee}$.
The portion of the band for each solar model below the HM bound is 
the viable region.  
The portion of each band above the GENIUS/EXO line will be 
probed by these experiments.
Some implications for $m_3$ and $E_R$ are evident.
For example, for the SMA or $\phi_{12}=0$ solutions,
$m_3$ is bounded from above by 0.27~eV, and $E_R$ is bounded 
from below by 15~ZeV due to the HM exclusion;
$m_3$ is also bounded from below by 
$\sqrt{\dmatm}\sim 0.05$~eV and $E_R$ from above by 80~ZeV. 
As $\phi_{12}$ increases from zero to $\pi$, the \znbb\ 
upper bound on $m_3$ increases to 1~eV and 3~eV, respectively,
for the 90\% C.L. LMA and LOW solutions;
the \znbb\ lower bound on the Z-burst energy decreases to
4~ZeV and 1~ZeV for these same LMA and LOW solutions, respectively.

On may turn the correlation between \znbb\ and the Z-burst model around.
As an example,
if the Z-burst energy $E_R$ is fixed, e.g. by 
assuming a factor 100 CNB density increase in the Galactic halo
due to clustering, one has $m_3\sim 0.5$~eV and $E_R\sim 8$~ZeV, 
and a resulting lower bound on $m_{ee}$ of 0.1~eV and 0.04~eV
in the LMA and LOW models at 90\% C.L., respectively.
As can be seen in Fig.\ 4, these values of $m_{ee}$ lie within 
the reach of the 1 ton GENIUS and the EXO proposals.
Therefore, if the GENIUS/EXO experiments fail to see \znbb,
then either neutrinos are Dirac particles, 
neutrino clustering in our Galactic halo is insignificant,
or the Z-burst hypothesis is wrong.
A more complete list of correlated inferences is now given.

If \znbb\ were to measure a value of $m_{ee}$ above 0.01~eV, then
the implications for the Z-burst model are:

\begin{itemize}

\item
The absolute mass $m_3$ and therefore the Z-burst energy 
$E_R$ will be determined with an 
accuracy factor of $(\cos 2 \theta_{\rm sun})^{-1}$, which is 
unity for the SMA solution but $\sim 4-10$ at present for large-mixing
solutions.

\item
If $m_1$ is shown to exceed $\sim 0.05$~eV,
then the three neutrino masses are near-degenerate,
and the absolute rate of Z-bursts is increased by three,
independent of the resonant neutrino flux.

\item
Depending on what absolute mass scale is discovered,
a factor of 100 (for $m_3\sim 0.5$~eV) to $10^3$
(for $m_3 \sim 1$~eV)  may be gained in the Z-burst rate
due to clustering in the Galactic halo.

\item
The neutrino is definitely a Majorana particle,
and so a factor of two more is gained in the Z-burst rate relative 
to the Dirac neutrino case;
this is because the two active helicity states of the relativistic CNB
depolarize upon cooling to populate all spin states (two active and
two sterile states for Dirac neutrinos, 
but only the original two active states for Majoranas) \cite{Ringberg}.

\end{itemize}

If \znbb\ will not be observed with $m_{ee}$ as low as $\sim 0.01$~eV,\\
then either:

\begin{itemize}

\item
The absolute mass $m_3$ is determined to be 
$m_3\simeq \sqrt{\dmatm}\sim 0.5$~eV; 
and no halo clustering and no mass-degeneracy enhance the Z-burst model.

\end{itemize}

\noindent or: 

\begin{itemize}

\item
neutrinos are Dirac particles.

\end{itemize}

Conversely, if
the Z-burst model turns out to be the correct explanation of EECRs, then
it is probable that neutrinos possess one or more masses in the range 
$m_{\nu}\sim (0.1-1)$~eV. Reference to Fig.\ 1 reveals that
mass-degenerate neutrino models are then likely. 
Some consequences are:

\begin{itemize}

\item
A value of $m_{ee}>0.01$~eV results, and thus
a signal of \znbb\ in the GENIUS/EXO experiments is predicted,
assuming the neutrinos are Majorana particles.

\item
Neutrino mass is sufficiently large to affect the 
CMB/LSS power spectrum. 

\end{itemize}

We have illustrated in some detail 
the correlation between \znbb\ and Z-bursts. The extension to
the tritium end-point experiment and the CMB/LSS study is 
straightforward.
At a minimum, the \znbb\ and tritium end-point experiments will 
cross-check each other over a significant range of $m_1$ 
(assuming of course, that neutrinos are Majoranas).
And the similarity in reach of the Z-burst and 
the CMB/LSS approaches allows a cross-check over a significant 
range of $m_3$; in particular, independent confirmation of the
existence of the CNB is available.
At a maximum, each of the four experimental approaches 
impacts the other three, since all three $m_j$'s are related by
the oscillation $\dmsq$'s.

\section{Discussion and conclusions}

The mass-squared differences inferred from oscillation 
interpretations of solar and atmospheric
neutrino data relate the three neutrino mass-eigenvalues
$m_3$, $m_2$, and $m_1$.  Accordingly, only the overall
mass scale is devoid of information.  We have considered
four approaches mixing experiment and theory 
which have the potential to infer a neutrino mass below 1~eV.  
These are \znbb\ and tritium decay
end-point measurements, which in future experiments
may be sensitive to the lighter masses
$m_1$ and $m_2$ if they exceed 0.01 and 0.2~eV, respectively;
and extreme-energy cosmic ray measurements 
in the context of the Z-burst model,
and comparisons of cosmological measurements of CMB fluctuations
and LSS distributions, which are sensitive to all neutrino masses
$\gsim 0.1$~eV.
Due to the mass relations implied by the oscillation data,
the findings expected from each of these four approaches
is correlated with the findings expected in the other three.
We have presented in some detail why and how this is so.
Special emphasis was placed on \znbb\ as representative of 
$m_1$, $m_2$ measurements, and on the Z-burst model as representative
of $m_3$ measurements.  The gross correlations between \znbb\ and the
Z-burst model are presented in Fig.\ 4.  More subtle inferences
are itemized in the previous section.
Present constraints from the tritium decay spectrum and from
CMB/LSS measurements are shown in Fig.\ 1.
Taken together, the four approaches  
hold the potential not only to determine the
absolute neutrino mass, but also to cross-check the validity of
the assumptions underlying the approaches.

Of the four approaches discussed here,
the Z-burst model probably contains the most speculative assumption,
namely that there exists a substantial cosmic flux of neutrinos
at energy $E_R\sim 10^{22}$~eV.  This assumption may be checked 
directly \cite{Ringberg} in a teraton neutrino detector
such as the proposed EUSO/OWL/AW orbiting experiment \cite{EUSO/OWL}.
The remaining assumptions in the Z-burst model seem solid, relying
only on Standard Model physics, the Standard Cosmological Model,
and the existence of neutrino mass. 
In other words, if $F_\nu (E_R)$ is nonzero,
then Z-bursts have to occur; but the rate is proportional
to the completely unknown value of $F_\nu (E_R)$.\footnote
{
It was proposed long ago \cite{wei82} that spectroscopy of all
neutrino masses may be done by observing the energies 
($E^R_{\nu_j}=4\,({\rm eV}/m_j)$~ZeV) of 
absorption dips in the extreme-energy neutrino flux.
This idea remains possible in principle, but in practice
requires an even larger neutrino flux than that 
required for an observable rate of Z-bursts.}

Finally, we wish to comment on what changes in this work 
if the neutrino masses exhibit the disfavored 
``inverted'' spectrum.  In the inverted spectrum, 
the two heavier states are split from one another by $\dmsun$, 
and separated from the remaining lighter state by $\dmatm$.
With the inverted spectrum, the ordinate Fig.\ 1 
becomes the near-degenerate masses 
of the two heavier states $m_3$ and $m_2$, while the abscissa
becomes just the single lighter state $m_1$.
More importantly,
the $\nue$ state is mixed mainly with the two heavier states,
and so the \znbb\ and tritium end-point approaches 
to absolute mass determination become sensitive to the heavier states,
as is the case with the Z-burst and CMB/LSS approaches. 
In the \znbb\ section of this paper, 
the lighter mass $m_1$ in eqns.\ (\ref{dbeqn2}) and
(\ref{dbbnds}) is replaced with the heavier mass $m_3$.
As a consequence, the present HM bound on $m_{ee}$ directly impacts 
the Z-burst model, and the potential of CMB/LSS measurements
to infer a neutrino mass. 
For the degenerate case with $m_1 \gg 0.05$~eV,
the situation is equivalent to the normal hierarchy.
However, even in the strongly (inverse) hierarchical case, where 
$m_1 \ll m_3$, it is true that $m_{ee}\gsim m_3 \cos 2 \theta_{\rm sun} \gsim
\sqrt{\dmatm}\cos 2 \theta_{\rm sun}$, which allows 
a cross-check of these approaches.

With four approaches available for determination of neutrino mass
below 1~eV, there is hope that the absolute neutrino mass scale
will become known.  The overlap in mass-reach 
of the four approaches, discussed in this work, 
will provide an important consistency check
on any positive result of any one approach.

\section*{Acknowledgements}
This work was supported in part by the
DOE grant no.\ DE-FG05-85ER40226.

\newpage
\begin{figure}
\epsfxsize=120mm
\epsfbox{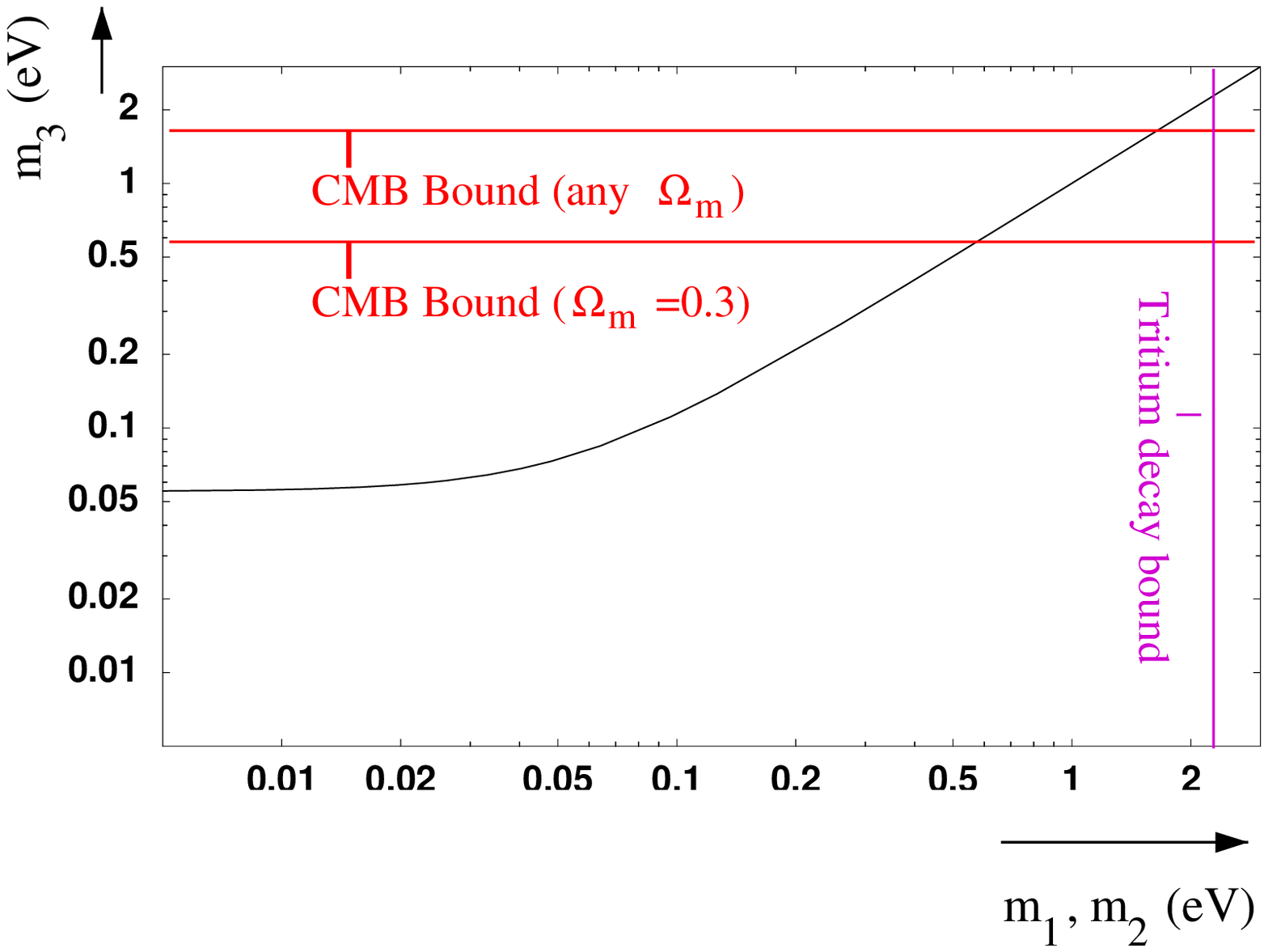}
\caption{ Neutrino mass constraints in the $m_{1,2}$-$m_3$
plane. The curved line corresponds to allowed values according to the solar
and atmospheric neutrino data. 
Direct mass measurements from CMB and tritium beta decay
exclude the regions beyond their respective straight lines.  
} 
\end{figure}

\newpage

\begin{figure}
\epsfxsize=120mm
\epsfbox{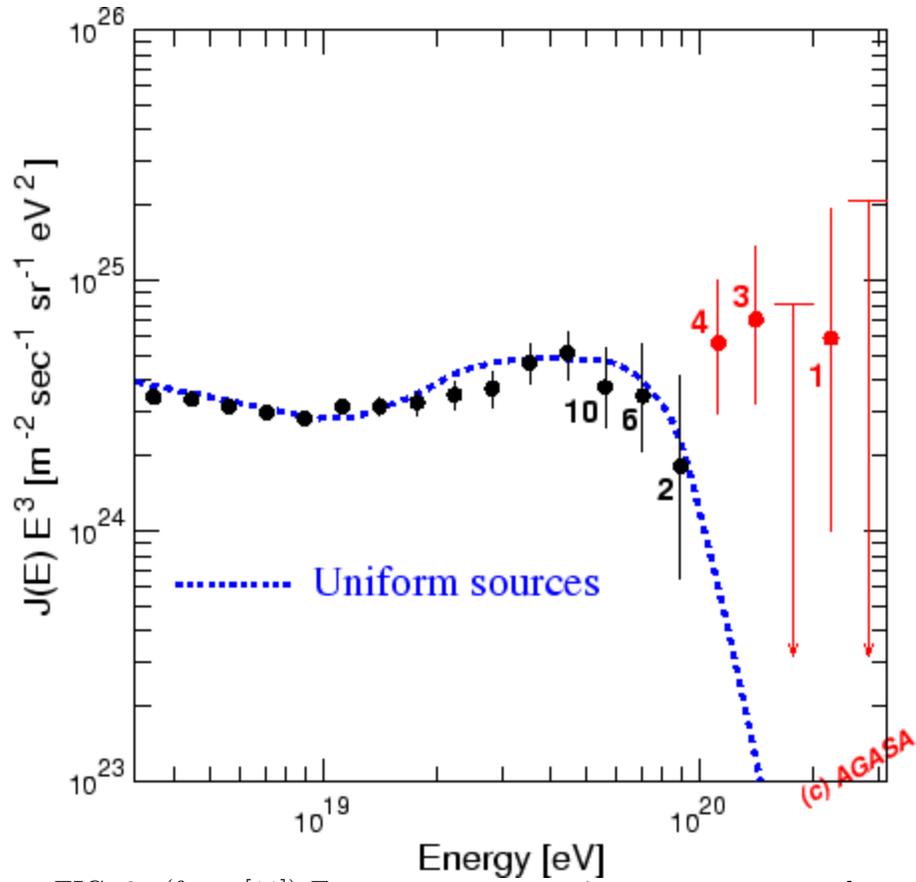}
\caption{(from {\protect \cite{agasa}}) Extreme-energy cosmic ray 
spectrum as observed by AGASA. Error bars correspond
to 68 \% C.L. and the numbers count the events per energy bin.
The dashed line revealing the GZK cutoff 
is the spectrum expected from uniformly distributed 
astrophysical sources.} 
\end{figure}

\newpage

\begin{figure}
\epsfxsize=120mm
\epsfbox{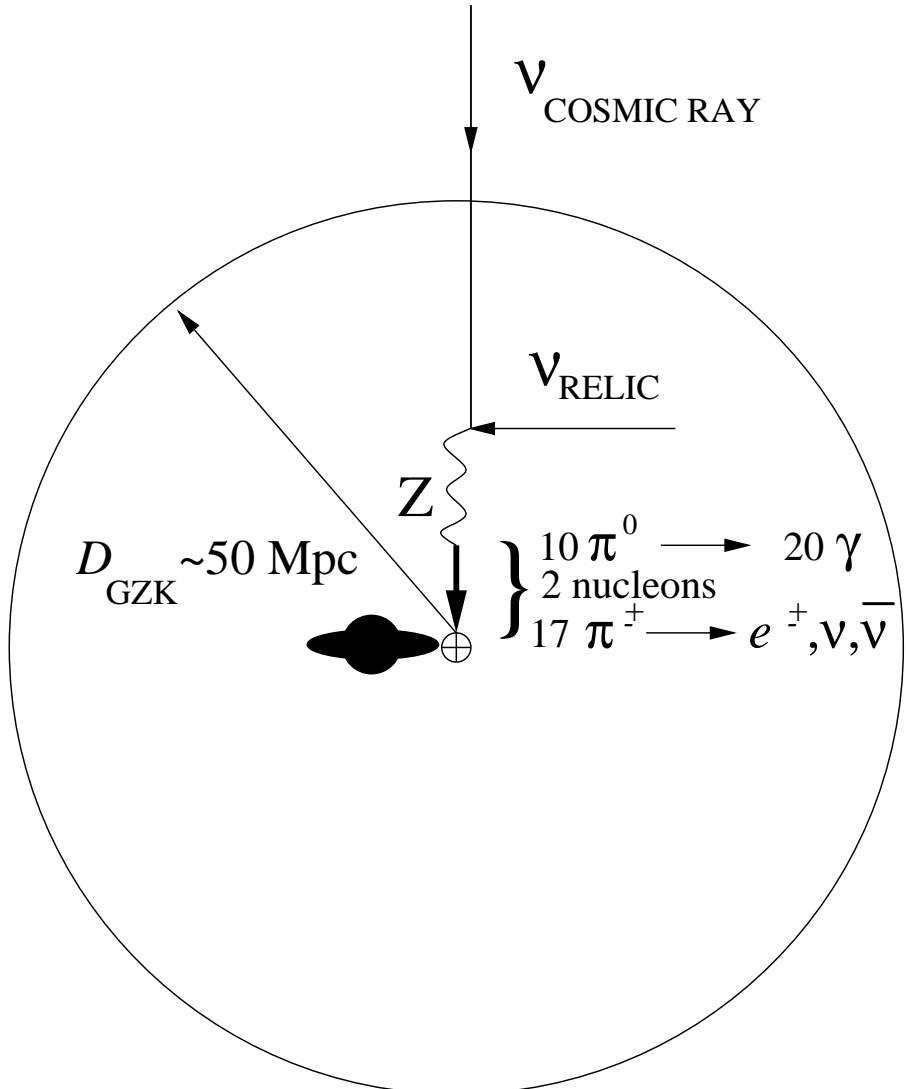}
\caption{
Schematic diagram showing the production of a Z-burst resulting from the 
resonant annihilation of a cosmic ray neutrino on a relic (anti)neutrino.
If the Z-burst occurs within the GZK zone ($\sim$ 50 to 100 Mpc) and is 
directed towards the earth, then photons and nucleons with energy above 
the GZK cutoff may arrive at earth and initiate super-GZK air-showers.
} 
\end{figure}

\newpage

\begin{figure}
\epsfxsize=120mm
\epsfbox{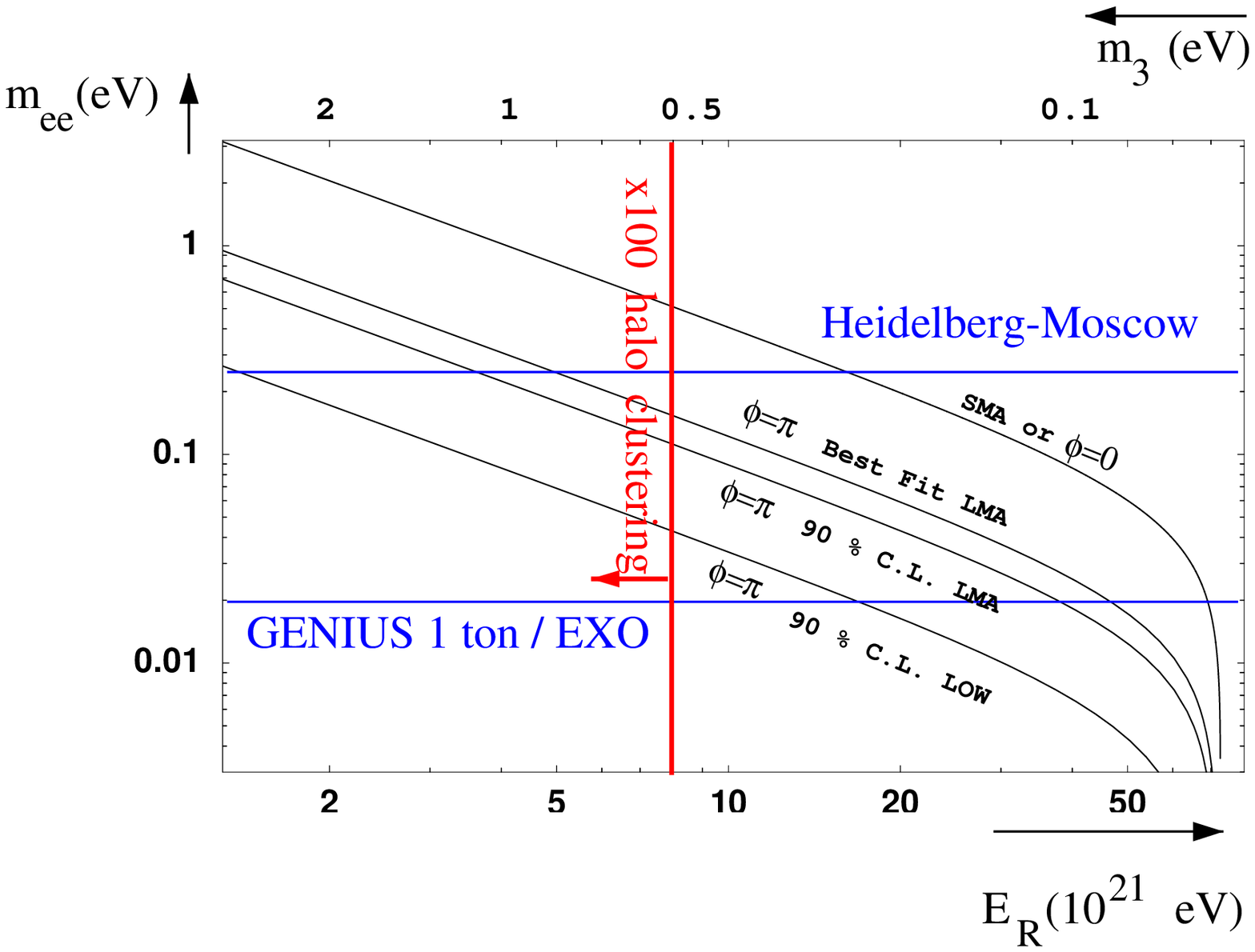}
\caption {No-neutrino double beta-decay observable $m_{ee}$ 
versus mass of the heaviest neutrino
$m_3$, or, alternatively, the resonant Z-burst energy $E_R$.
The curved lines show allowed regions for different solutions of the solar 
neutrino anomaly; from top to bottom, the case for $\phi_{12}=0$, or arbitrary 
$\phi_{12}$ for the SMA, and the cases $\phi_{12}=\pi$ for the LMA solution
best fit, the 90 \% C.L. LMA solution and the 90 \% C.L. LOW solution.
The region between the $\phi_{12}=0$ 
and the $\phi_{12}=\pi$ lines are allowed in the various solar 
solutions. The straight lines show the bound from the Heidelberg--Moscow 
experiment excluding the region above, the sensitivity of the 1 ton GENIUS and
EXO projects, and the region where Galactic halo clustering provides a 
neutrino overdensity of 100 or more (moving to the left).
} 
\end{figure}


\begin{thebibliography}{200}

\bibitem{bww98}
V. Barger, T.J. Weiler, K. Whisnant, Phys.Lett. B442 (1998) 255-258

\bibitem{tritium}
C. Weinheimer, talk at the EPS HEP 99 conference at Tampere/Finland

\bibitem{primack00}
R. Croft, W. Hu, and R. Dave, Phys. Rev. Lett. 83, 1092 (1999);
M. Fukugita, G.-C. Liu, and N. Sugiyama, Phys. Rev Lett. 84, 1082 (2000);
for a review, see, e.g., 
C-P Ma, TASI98 Lecture, Boulder, CO, 
ed.\ P. Langacker, Pub.\ World Sci.\ [astro-ph9904001];
J. Primack and M. Gross, in ``Current Aspects of Neutrino Physics'', 
ed.\ D. Caldwell, pub.\ Springer-Verlag, 2001 [astro-ph/0007165].

\bibitem{bbdata}
H.V. Klapdor-Kleingrothaus et al., to be publ. 2000 and
http://www.mpi-hd.mpg.de/non\_acc/main.html

\bibitem{numassrevs} 
R. Mohapatra, in ``Current Aspects of Neutrino Physics'', 
ed.\ D. Caldwell, pub.\ Springer-Verlag, 2001 [hep-ph/9910365].
G. Altarelli and F.Feruglio, hep-ph/9905536, to appear in Phys. Rept.


\bibitem{Smirnov98}
Theories with large extra dimensions can motivate small neutrino masses
of either Dirac or Majorana type. 
Examples for Dirac masses are:\\
N. Arkani-Hamed, S. Dimopoulos, G. Dvali, J. March-Russell,
hep-ph/9811448;
K. Dienes, E. Dudas, and T. Gherghetta, Nucl. Phys. B557, 25 (1999);
A. Faraggi and M. Pospelov, Phys. Lett. B458, 237 (1999);
A. Das and O. Kong, Phys. Lett. B470, 149 (1999);
G. Dvali, A. Yu. Smirnov, Nucl.Phys. B563 (1999) 63;
R.N. Mohapatra, S. Nandi, A. Perez-Lorenzana, Phys.Lett. B466 (1999) 115;
A. Ioanissian, J.W.F. Valle, hep-ph/9911349;
R. Barbieri, P. Creminalli, and A. Strumia, hep-ph/0002199.


\bibitem{MAP}The homepage for the Microwave Anisotropy Probe (MAP) is\\ 
http://map.gsfc.nasa.gov/

\bibitem{Planck} The homepage for the PLANCK Mission is \\
http://astro.estec.esa.nl/SA-general/Projects/Planck/

\bibitem{SDSS} The Sloan Digital Sky Survey (SDSS) homepage is\\ 
http://ssds.nasa.gov/

\bibitem{2dF} The Two Degree Field system (2dF) homepage is \\
http://www.ast.cam.ac.uk./AA0/2df/

\bibitem{agasa}
The homepage for the Akeno Giant Air Shower Array (AGASA) is\\
http://www-akeno.icrr.u-tokyo.ac.jp/AGASA/

\bibitem{hires}
The homepage for the High Resolution Fly's Eye Cosmic Ray Detector  
(HiRes) is\\
http://hires.physics.utah.edu/

\bibitem{auger}
The homepage for the Pierre Auger Project is\\
http://www.auger.org/

\bibitem{telarr}
The homepage for the Telescope Array Project is\\
http://www-ta.icrr.u-tokyo.ac.jp/

\bibitem{EUSO/OWL} The Extreme-Universe Space Observatory (EUSO) homepage is\\
http://www.ifcai.pa.cnr.it/Ifcai/euso.html;\\
the Orbiting Wide-angle Light-collectors (OWL) homepage is\\ 
http://owl.gsfc.nasa.gov;\\ the AirWatch side of OWL has a homepage at\\
http://www.ifcai.pa.cnr.it/$\sim$AirWatch/


\bibitem{wei82} T.J. Weiler, Phys. Rev. Lett. (1982); 
   ibid., Astrophys. J. 285, 495 (1984); 
E. Roulet, Phys. Rev. D47, 5247 (1993).

\bibitem{Zburst}
T.J. Weiler, Astropart. Phys. 11, 303 (1999),
and ibid.\ 12, 379E (2000) [for corrected receipt date];
D. Fargion, B. Mele and A. Salis, Astrophys. J. 517, 725 (1999).

\bibitem{BBM00}
J. Beacom, R. Boyd, and A. Mezzacappa, Phys. Rev. Lett. 85, 3568 (2000).

\bibitem{inv} 
H. Minakata and H. Nunokawa, hep-ph/0010240;
C. Lunardini and A. Yu. Smirnov, hep-ph/0009356.

\bibitem{gon99}
M.C. Gonzalez-Garcia, M. Maltoni, C. Pe\~na-Garay, 
J.W.F. Valle, hep-ph/0009350;
For earlier works see J.N. Bahcall, P.I. Krastev, A.Yu. Smirnov,
Phys.Rev. D60 (1999) 093001;
G.L. Fogli, E. Lisi, D. Montanino, A. Palazzo, Phys.Rev. D62 (2000) 013002;
G.L. Fogli, E. Lisi, D. Montanino, A. Palazzo, Phys.Rev. D62 (2000) 113004;
H. Minakata, H. Nunokawa, hep-ph/9902460;
see also M.V. Garzelli and C. Giunti, hep-ph/0012044.

\bibitem{SN87a} M. Kachelriess, R. Tomas, and J. Valle, hep-ph/0012134,
    and references therein.

\bibitem{EECRdata}
M. Takeda et al. (AGASA Collaboration),
   Phys. Rev. Lett. 81, 1163 (1998) [astro-ph/9807193];
D.J.Bird et al. (Fly's Eye Collaboration), Phys. Rev. Lett. 71, 3401 (1993),
   Astrophys. J. 424, 491 (1994), and ibid 441, 144 (1995);
M.A. Lawrence, R.J.O.Reid and A.A. Watson (Haverah Park Collaboration),
     J. Phys. G 17,773 (1991),
and M. Ave, J.A.Hinton, R.A.Vazquez, A.A.Watson and E. Zas,
    Phys. Rev. Lett. 85, 2244 (2000),
and http://ast.leeds.ac.uk/haverah/hav-home.shtml;
N.N. Efimov et al. (Yakutsk Collaboration),
    Proc. ``Astrophysical Aspect of the Most Energetic Cosmic Rays,''
    p. 20, eds. M. Nagano and F. Takahara, World Sci., Singapore, 1991;
D. Kieda et al. (HiRes Collaboration), Proc.\ of the 26th ICRC, Salt Lake
City, Utah, 1999.

\bibitem{EECRreviews}
Recent reviews include:
P. Biermann, J. Phys. G23, 1 (1997);
P. Bhattacharjee and G. Sigl, Phys. Rept. 327, 109 (2000) [astro-ph/9811011];
R.D. Blandford, in ``Particle Physics and the Universe'',
Physica Scripta, ed. L. Bergstrom et al.,  World Scientific [astro-ph/9906026];
A.V. Olinto, ``David Schramm Memorial Volume'' of Phys. Rept. 333, 329 (2000);
X. Bertou, M. Boratov, and A. Letessier-Selvon, Int. J. Mod. Phys. A15,
2182 (2000);
A. Letessier-Selvon, Lectures at ``XXVIII International Meeting on
Fundamental Physics'',
Cadiz, Spain (2000) [astro-ph/0006111];
M. Nagano and A.A. Watson, Rev. Mod. Phys. 72, 689 (2000);
F.W. Stecker, astro-ph/0101072.
%

\bibitem{RPP}
Particle Data Group, Euro. Phys. J. C15 (2000) pp. 226-7.

\bibitem{nuFlux}
S. Yoshida, G. Sigl, and S. Lee, Phys. Rev. Lett. 81, 5505 (1998);
J. Blanco-Pillado, R. Vazquez, and E. Zas, Phys. Rev. D61, 123003 (2000);
G. Gelmini and A. Kusenko, Phys. Rev. Lett. 82, 5202 (1999); 
ibid, 84, 1378 (2000);
J. Crooks, J. Dunn, and P. Frampton, astro-ph/0002089.

\bibitem{TG80s} 
S. Tremaine and J. Gunn, Phys. Rev. Lett. 42 (1979) 407 

%
%

\bibitem{tegmark}
W. Hu, D.J. Eisenstein, and M. Tegmark, 
Phys. Rev. Lett. 80 (1998) 5255.

\bibitem{lopez}
R.E. Lopez, astro-ph/9909414.


\bibitem{kla00}
H.V. Klapdor-Kleingrothaus, H. P\"as, A.Y. Smirnov, hep-ph/0003219, 
Phys. Rev D, in the press;
M. Czakon, J. Gluza, J. Studnik, M. Zralek, hep-ph/0010077,
and references therein. 


\bibitem{genius}
J. Hellmig, H.V. Klapdor--Kleingrothaus, Z. Phys. A 359 (1997) 351;
H.V. Klapdor--Kleingrothaus, M. Hirsch, Z. Phys. A 359 (1997) 361;
H. V. Klapdor-Kleingrothaus, L. Baudis, G. Heusser, B. Majorovits, H. P\"as,
hep-ph/9910205. 

\bibitem{exo}
M. Danilov et al., Phys.Lett. B480 (2000) 12.

\bibitem{chooz}
M. Apollonio et al. (CHOOZ collab.), hep-ex/9907037, 
Phys. Lett. B 466 (1999) 415.

\bibitem{barger00}
V. Barger, D. Marfatia, B.P. Wood, hep-ph/0011251;
H. Murayama and A. Pierce, hep-ph/0012075.

\bibitem{Ringberg} T.J. Weiler, 
    in ``Beyond the Desert 99, Ringberg Castle'', Tegernsee, Germany, 
    June 6-12, 1999 [hep-ph/9910316].


\end{thebibliography}
\end{document}